\documentclass{article}
\usepackage{graphicx}
\usepackage{amsmath}
\usepackage{amsfonts}
\usepackage{amssymb}

\begin{document}

\title{Implementation of a $CNOT$ gate in two cold Rydberg atoms by the nonholonomic
control technique }
\author{E. Brion$^{1}$, D. Comparat$^{1}$ and G. Harel$^{2}$ 
\\$^{1}$\emph{Laboratoire Aim\'{e} Cotton, CNRS\ II}, \\ \emph{B\^{a}timent 505, 91405 Orsay Cedex, France.}
\\$^{2}$\emph{Department of Computing, University of Bradford}, \\ \emph{Bradford, West Yorkshire BD7 1DP, United Kingdom.}}
\maketitle
\begin{abstract}
We present a demonstrative application of the nonholonomic control method to a
real physical system composed of two cold Cesium atoms. In particular, we show
how to implement a $CNOT$ quantum gate in this system by means of a controlled
Stark field. 
\end{abstract}

\section{Introduction}

Quantum control is a very topical issue which bears relevance to many different fields of
contemporary physics and chemistry, such as Molecular Dynamics in laser fields
and Quantum Optics \cite{TR85,PK02,SB86,PDR88,PMZK93,LE96}. A few examples of control of the quantum state by adiabatic transport \cite{PMZK93}, by unitary evolution \cite{LE96} or by conditional
measurements \cite{VAS93,HKIC96}
have been already proposed for the particular quantum system
of atoms interacting with quantized electromagnetic field in a single-mode
resonator. 

In parallel, a theoretical framework of quantum control has been
built up. Four different types of problems have
been identified in the literature \cite{BS90,SPS03}: the control of pure state, the control of
density matrix, the control of observable and, finally, the control of the
evolution operator. Each of these problems can be formulated in the same way: the goal is to impose on the considered characteristics an arbitrarily chosen value.  

To achieve a control objective, one has to perturb the system, since its
natural evolution usually results in too restrictive a dynamics. The control
Hamiltonian $\widehat{H}\left(  t\right)  $ thus comprises the unperturbed
Hamiltonian $\widehat{H}_{0}$ as well as $M$ Hamiltonians of the form
$C_{m}\left(  t\right)  \widehat{P}_{m}$, which stand for the interaction
Hamiltonians of the system with $M$ classical fields
of controllable amplitudes $C_{m}\left(  t\right)  $,
\[
\widehat{H}\left(  t\right)  =\widehat{H}_{0}+\sum C_{m}\left(  t\right)
\widehat{P}_{m}.
\]
The functions $\left\{  C_{m}\left(  t\right)  \right\}  $ play the role of
the control parameters one has to adjust in order to achieve the desired
control process. Any problem of control can thus be put in the following form:
for a given physical system, specified by $\widehat{H}_{0}$ and
$\left\{ \widehat{P}_{m} \right\}$,
find the values of the control parameters $\left\{  C_{m}\left(  t\right)  \right\}  $
which ensure that a specific characteristics of the system
(quantum state, density matrix, observable, evolution operator)
will take an arbitrarily prescribed value.

Not all the objectives are always feasible.
For example, the unitarity of the evolution operator for closed systems prevents the eigenvalues of the density matrix from changing through a Hamiltonian
process of control. This kind of constraints is often referred to as
\emph{kinematical constraints} \cite{SSL02}. But there also exist \emph{dynamical
constraints} which stem from the algebraic properties of the Hamiltonians
$\left\{ \widehat{H}_{0},
\widehat{P}_{m}\right\}  $. Indeed, the evolution operator
\[
\widehat{U}\left(  t\right)  =\mathcal{T}\left\{  e^{-\frac{i}{\hbar}\int
_{0}^{t}\widehat{H}\left(  \tau\right)  d\tau}\right\}
\]
where $\mathcal{T}$ denotes the chronological product, belongs to the Lie
group obtained by the exponentiation of the Lie algebra generated by the operators
$\left\{  i\widehat{H}_{0},i\widehat{P}_{m}\right\}  $. The feasibility of a particular problem of control in a specific physical
situation, defined by the Hamiltonians
$\left\{ \widehat{H}_{0}, \widehat{P}_{m}\right\}  $, is
clearly related to the properties of this algebra. For example, if
one wants to completely control the evolution operator of a quantum system,
\emph{i.e.}
to be able to give the operator $\widehat{U}$ any prescribed value,
one must perturb the system in such a way that the operators
$\left\{  i\widehat{H}_{0},i\widehat{P}_{m}\right\}  $ generate the whole Lie
algebra $u\left(  N\right)  $ which provides, through exponentiation, the
whole Lie group $U\left(  N\right)  $ \cite{JS72,RSDRP95} (this prescription is called the Bracket Generation Condition). Necessary mathematical conditions also
exist for the other types of control problems which can be found in
the literature \cite{SPS03}: these
conditions are obviously weaker than the previous one,
since the controllability of the evolution operator automatically implies all the other ones.

The feasibility of a control problem can thus be decided through mathematical
criteria established in the context of the Lie group theory. But the explicit
values of the control parameters achieving the desired control objective still
remain to be found. In other words, once the existence of a solution has been
proved, one still has to find it explicitly. To achieve this goal, different methods, such as optimal control \cite{OKF98,SGL00,PK02}, have been proposed, most of which rely either on a known or intuitively guessed particular solution which can be further optimized with respect to a given cost functional, through variational schemes \cite{BS90}. A purely algebraic approach \cite{SGRR01}, based on
the decomposition of the arbitrary desired evolution on the Lie group, is also
possible, but rapidly leads to intractable computations as the dimension of
the state space increases.

In the context of the control of the evolution operator, a constructive method called nonholonomic control \cite{HA99} was proposed, in the same spirit as in \cite{SL95}:
this method is fundamentally algebraic but also uses optimization steps.
The physical idea is to alternately apply two distinct well chosen perturbations $\widehat{P}_{a}$ and $\widehat{P}_{b}$ (\emph{i.e.} two perturbations which check the Bracket Generation Condition). The timings of the interaction pulses play the role of control parameters and are determined by solving the \textquotedblright inverse Floquet problem\textquotedblright.
The convergence of the algorithm results from an unsuspected simplification which emerges from the Random Matrix Theory. Indeed, it relies on the algebraic properties of the $N^{th}$ roots of the identity matrix, the spectra of which resemble those of random unitary matrices which are ruled by the Dyson distribution law.
This method, combined with a generalization of the Quantum Zeno effect,
has also led to coherence protection schemes \cite{BACDHKKMP05,BADHK05}. 

The nonholonomic technique is completely universal and can thus be applied to any physical system.
In the last few years, cold atomic Rydberg states have appeared as particularly relevant in the context of quantum information, and have been widely investigated both theoretically \cite{JCZRCL00,LFCDJCZ01} and experimentally \cite{TFSKZCEG04,SRAMW04}.
In this paper we propose to consider a real system composed of two cold
Cesium (Cs)
atoms interacting via dipolar forces and prepared in Rydberg states. Under some physical assumptions, the Hilbert space of the system is restricted to four states, as for a two qubit system. We then propose a control experiment employing a pulsed electric field which allows us to impose a $CNOT$ gate to the system through nonholonomic control. Even though we have tried to propose a realistic experimental setting, the simplifications we have made result in serious limitations of the scheme we present.
Nevertheless, it shows how the nonholonomic method can work on a real physical
system and suggests that this technique can be an effective way to solve
real-life problems of control.     

The paper is organized as follows. In Sec.II, we recall briefly the main features of the nonholonomic control technique. In Sec.III, we describe our physical application in detail - the system, the control Hamiltonians, the calculated control parameters - and discuss its validity and limitations.

\section{Control of the evolution through nonholonomic control}

Let us consider an $N$-dimensional quantum system of unperturbed Hamiltonian
$\widehat{H}_{0}$. Our goal is to control its evolution operator $\widehat{U}%
$, \emph{i.e.} to be able to achieve any arbitrary evolution $\widehat
{U}_{arbitrary}\in U\left(  N\right)  $.\
\begin{figure}
[ptb]
\begin{center}
\includegraphics[
height=1.9199in,
width=2.7717in
]%
{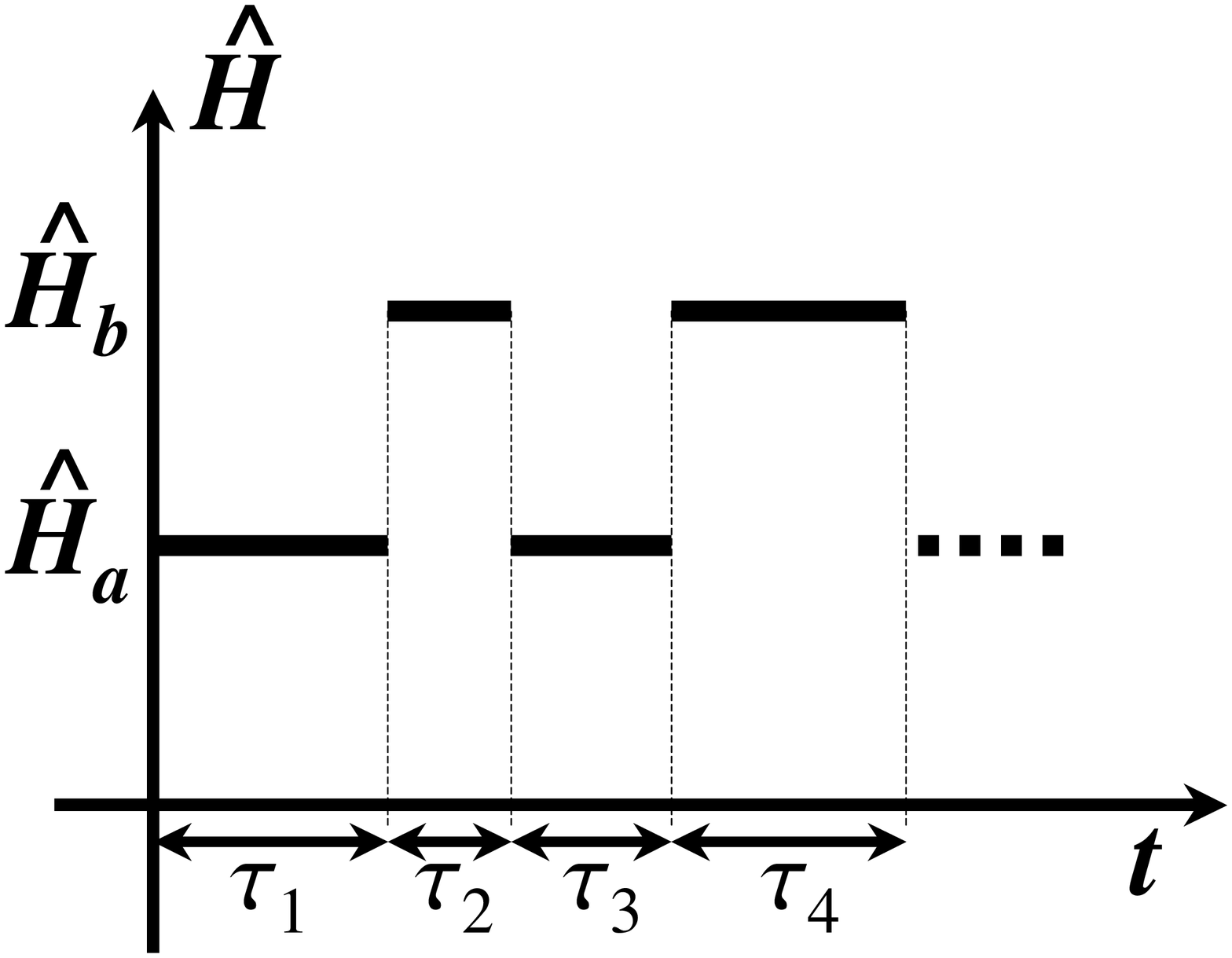}%
\caption{Pulsed shape of the control Hamiltonian.}%
\label{Fig1}%
\end{center}
\end{figure}
To this end, we alternately apply two physical perturbations, of Hamiltonians
$\widehat{P}_{a}$ and $\widehat{P}_{b}$, during $N^{2}$ time intervals
(pulses), the timings of
which are denoted by $\left\{  \tau_{k}\equiv t_{k}-t_{k-1}\right\}
_{k=1,\ldots,N^{2}}$ ($t_{0}=0$ and $t_{N%
%TCIMACRO{\U{b2}}%
%BeginExpansion
{{}^2}%
%EndExpansion
}=T$ correspond to the beginning and the end of the control sequence,
respectively). The control Hamiltonian takes the following pulsed shape (cf
Fig. \ref{Fig1}):%
\[
\widehat{H}(t)=\widehat{H}_{0}+C_{a}\left(  t\right)  \widehat{P}_{a}%
+C_{b}\left(  t\right)  \widehat{P}_{b}%
\]
with $C_{a}\left(  t\right)  =1$,  $C_{b}\left(  t\right)  =0$ and
$\widehat{H}(t)=\widehat{H}_{0}+\widehat{P}_{a}\equiv\widehat{H}_{a}$ for $t\in\left[  t_{2k},t_{2k+1}\right]$
and $C_{a}\left(  t\right)   =0$, $C_{b}\left(  t\right)  =1,$ and
$\widehat{H}(t)=\widehat{H}_{0}+\widehat{P}_{b}\equiv\widehat{H}_{b}$ for $t\in\left[  t_{2k-1},t_{2k}\right],$
where $k=1,\ldots,N^{2}$.
The total evolution operator is therefore
\[
\widehat{U}\left(  \left\{  \tau_{1},\ldots,\tau_{N^{2}-1},\tau_{N^{2}}\right\}  \right) = e^{-\frac{i}{\hbar}\widehat{H}_{b}\tau_{N^{2}}}\cdot
e^{-\frac{i}{\hbar}\widehat{H}_{a}\tau_{N^{2}-1}} \ldots e^{-\frac{i}{\hbar}\widehat{H}_{a}\tau_{1}},
\]
where we have implicitly assumed that $N$ is even.

Our control problem can thus be translated into the following form: given an arbitrary unitary operator $\widehat{U}_{arbitrary}\in U(N),$ we want to find a time vector $\overrightarrow{\tau}=\left(\tau_{1},\ldots,\tau_{N^{2}}\right)$ with non-negative entries, such that
\begin{equation}
\widehat{U}\left(  \overrightarrow{\tau}\right)  =\widehat{U}_{arbitrary}.
\label{pbcont}%
\end{equation}

As we said previously, for a solution to exist the operators $\left\{
i\widehat{H}_{a},i\widehat{H}_{b}\right\}  $ must generate the whole Lie
algebra $u\left(  N\right)  $. This prescription can be checked directly as long
as the dimension $N$ is not too big: one simply computes the commutators of
all orders of $i\widehat{H}_{a}$ and $i\widehat{H}_{b}$ and stops as soon as
they generate $u\left(  N\right)  $. But when $N$ becomes large, direct
computation is intractable. In that case, one can simply check the following
sufficient conditions, suggested by V.G. Kac \cite{HA99},
according to which the system
becomes nonholonomic, that is completely controllable, when the
representative matrix of $\widehat{H}_{b}$ in the eigenbasis of $\widehat
{H}_{a}$ has no zero elements and vice versa, and also the eigenvalues and
their pairwise differences are distinct for both matrices.

Once the previous criterion is checked, one has to compute the time vector
$\overrightarrow{\tau}$ solution of Eq.(\ref{pbcont}). The method consists in determining the time vector $\overrightarrow{\tau}^{(0)}$ such that $\widehat{U}\left( \overrightarrow{\tau}^{(0)}\right)  =\widehat{I},
$ then iteratively approaching the time vector $\overrightarrow{\tau}$
through a Newton-like technique.

The straightforward way to compute $\overrightarrow{\tau}^{(0)}$\ would be to
minimize the functional $F\left(  \overrightarrow{\tau}\right)  =\Vert\widehat{U}\left(
\overrightarrow{\tau}\right)  -\widehat{I}\Vert^{2}$ with respect to $\overrightarrow{\tau}$. However, $F$ presents many local
minima which make its optimization uneasy. Yet, there exists an alternative
method based on the algebraic properties of the $N^{th}$ roots of the identity matrix. 

The idea is to look for $N$ parameters $\{T_{k}\}_{k=1...N}$ such that
\begin{equation}
e^{-\frac{i}{\hbar}\widehat{H}_{b}T_{N}}\cdot e^{-\frac{i}{\hbar}\widehat
{H}_{a}T_{N-1}}\ldots e^{-\frac{i}{\hbar}\widehat{H}_{a}T_{1}}=\widehat{I}%
^{\frac{1}{N}},
\label{RacNId}
\end{equation}
where $\widehat{I}^{\frac{1}{N}}$ is a non-degenerate $N^{th}$ root of the identity matrix,
\emph{i.e.} a matrix of the form
\[
\widehat{I}^{\frac{1}{N}}=\widehat{M}^{-1}\cdot\left[
\begin{array}
[c]{cccc}%
1 & 0 & \cdots & 0\\
0 & e^{i\frac{2\pi}{N}} & \cdots & 0\\
\vdots & \vdots & \ddots & \vdots\\
0 & 0 & \cdots & e^{i\left(  N-1\right)  \frac{2\pi}{N}}%
\end{array}
\right]  \cdot\widehat{M},
\]
where $\widehat{M}$ is a unitary matrix. To compute the $T_{k}$'s, we use the
following algebraic property: if $P_{\widehat{U}}(\lambda)\equiv\sum
_{j=0}^{N}a_{j}\lambda^{j}$ denotes the characteristic polynomial of a unitary
matrix $\widehat{U}$, then $\sum_{j=0}^{N}\left\vert a_{j}\right\vert ^{2}%
\geq2$ and the equality is achieved if and only if $\widehat{U}$\ is an $N^{th}$ root of
the identity matrix, up to a global phase factor. 

\begin{figure}[ptbh]
\begin{center}
\includegraphics[
width=2.5 in,
]{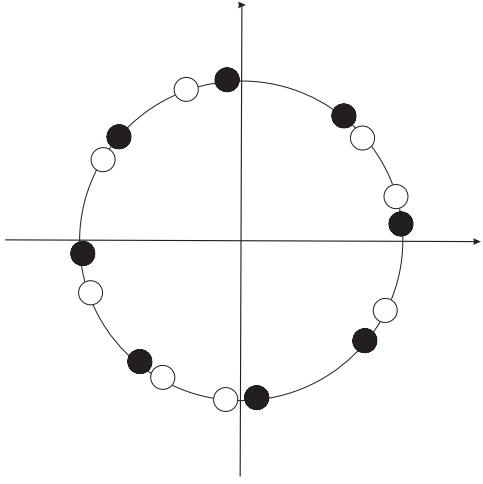}
\end{center}
\caption{Spectrum of a random unitary matrix (white circles) resulting from
the repulsion of the eigenvalues on a unit circle is shown vs the eigenvectors
of $N^{th}$ root of the identity matrix (black circles) multiplied by an
unimportant phase factor.}%
\label{Fig2}
\end{figure}

To obtain the $T_{k}$'s, one
thus computes the characteristic polynomial
\[ 
P_{\widehat{U}}(\lambda)\equiv\sum_{j=0}%
^{N}a_{j}\left(  \{T_{k}\}_{k=1...N}\right)  \lambda^{j}
\]
of the matrix product
\[
\widehat{U}=
e^{-\frac{i}{\hbar}\widehat{H}_{b}T_{N}}\cdot e^{-\frac{i}{\hbar}\widehat
{H}_{a}T_{N-1}}\ldots e^{-\frac{i}{\hbar}\widehat{H}_{a}T_{1}},
\]
and minimizes the function
\[
F_{N}=\sum_{j=0}^{N}\left\vert a_{j}\left(
\{T_{k}\}_{k=1...N}\right)  \right\vert ^{2}
\]
to $2$ with respect to the
$T_{k}$'s. This minimization turns to be quite easy, due to the fact that a
generic unitary matrix is very close to an $N^{th}$ root of the identity.\ In
fact, numerical work shows that in about 30\% cases of randomly chosen timings
$\left\{  T_{k}\right\}  $ the standard steepest descent algorithm immediately
finds the global minimum $F_{N}=2$. This fact has roots in the Random Matrix
Theory \cite{VA05}. Indeed, according to Dyson's law, the eigenvalues of random unitary
matrices tend to repel each other, and are thus very likely to be almost
regularly distributed on the unit circle, as those of an $N^{th}$ root of the
identity, as shown in Fig.\ref{Fig2}. In other words, in the space of
$N\times N$ unitary matrices, the $\widehat{I}^{\frac{1}{N}}$ matrices are present
in abundance, and can be reached from
a randomly chosen point by small variation
of the timings.

Finally, we define the time vector $\overrightarrow{\tau}^{(0)}$ corresponding
to the identity matrix by simple repetition of $\left\{  T_{k}\right\}  $
\begin{equation}
\tau_{i+\left(  j-1\right)  N}^{(0)}=T_{i}\mbox{ \ for \ }i,j=1,\ldots,N,
\label{temps}%
\end{equation}
and checks that indeed
\begin{eqnarray*}
\widehat{U}\left(  \overrightarrow{\tau}^{(0)} \right)   & = & \left( e^{-\frac
{i}{\hbar}\widehat{H}_{b}T_{N}}\cdot e^{-\frac{i}{\hbar}\widehat{H}_{a}T_{N-1}} \ldots e^{-\frac{i}{\hbar}\widehat{H}_{a}T_{1}} \right) \\
& \ldots & \left( e^{-\frac{i}{\hbar}\widehat{H}_{b}T_{N}}\cdot
e^{-\frac{i}{\hbar}\widehat{H}_{a}T_{N-1}}\ldots e^{-\frac{i}{\hbar}\widehat{H}_{a}T_{1}} \right) \\
& = & \left(\widehat{I}^{\frac{1}{N}} \right)^{N} = \widehat{I}
\end{eqnarray*}
up to an irrelevant global phase factor.

We now have to iteratively determine the time vector $\overrightarrow{\tau}$
from $\overrightarrow{\tau}^{(0)}$. Let us first consider the case of a target evolution close to the identity: in that case, $\widehat{U}_{arbitrary}$ can be
written in the form 
\begin{equation}
\widehat{U}_{arbitrary} = \widehat{U}_{\epsilon}\equiv\exp(-i\widehat{\mathcal{H}}\epsilon),
\end{equation}
where $\widehat{\mathcal{H}}$ is an $N \times N$ bounded ($||\widehat{\mathcal{H}}||\leq1$) dimensionless Hermitian Hamiltonian, and $\epsilon>0$ a small parameter.
We then calculate the variations $\delta \tau_{k}$, which are determined to
first order in $\epsilon$ by the linear equations
\begin{equation}
\sum_{k=1}^{N^{2}}\frac{\partial\widehat{U}}{\partial \tau_{k}} \left( \overrightarrow{\tau}^{(0)} \right)\,\delta
\tau_{k}=-i\widehat{\mathcal{H}}\epsilon. \label{eqlin}
\end{equation}
Once $\delta\overrightarrow{\tau}=
 \left(\delta\tau_{1},\ldots,\delta\tau_{N^{2}}\right)$
has been calculated through standard
techniques of linear algebra, we replace $\overrightarrow{\tau}^{(0)}$\ by
$\overrightarrow{\tau}^{(0)}+\delta\overrightarrow{\tau}$ and repeat the same
operation until we obtain $\overrightarrow{\tau}$ which checks $\widehat
{U}\left(  \overrightarrow{\tau}\right)  =\widehat{U}_{arbitrary}$ at the
desired accuracy.

If the evolution $\widehat{U}_{arbitrary}=\widehat{U}_{\epsilon}$ is not close to the identity,
that is, if $\epsilon$ is not small, one has to divide the work into elementary paths on which the previous method
converges. To this end, we consider an integer $n\geq2$ such that $\left(  \widehat
{U}_{arbitrary}\right)  ^{\frac{1}{n}} = \widehat{U}_{\frac{\epsilon}{n}}$ is attainable from $\widehat{I}$ through our iterative algorithm, and determine in this way the associated time vector
$\overrightarrow{\tau}_{\left(  \frac{1}{n}\right)  }$\ which checks
\[
\widehat{U}\left(  \overrightarrow{\tau}_{\left(  \frac{1}{n}\right)
}\right)  =\widehat{U}_{\frac{\epsilon}{n}}.
\]
Taking $\left(  \widehat{U}_{arbitrary}\right)  ^{\frac{1}{n-1}} = \widehat{U}_{\frac{\epsilon}{n-1}}$ as our new
target, we repeat the same algorithm to compute $\overrightarrow{\tau}_{\left(
\frac{1}{n-1}\right)  }$ such that
\[
\widehat{U}\left(  \overrightarrow{\tau}_{\left(  \frac{1}{n-1}\right)
}\right)  =\widehat{U}_{\frac{\epsilon}{n-1}},
\]
and so on.
We progress in this way as long as the algorithm converges: in
general, it stops at a value $n^{\ast}\geq1$, for which
Eq.(\ref{eqlin}) has no solution. Then, we keep the time vector
$\overrightarrow{\tau}_{\left(  \frac{1}{n^{\ast}}\right)  }$ and simply
repeat the same control sequence $n^{\ast}$ times to achieve the desired
evolution
\begin{eqnarray*}
\widehat{U}\left(\overrightarrow{\tau}_{\left(  \frac{1}{n^{\ast}}\right)  }\right) \ldots \widehat{U}\left(  \overrightarrow{\tau}_{\left(
\frac{1}{n^{\ast}}\right)  }\right)  & = &  \left( \widehat{U}_{\frac{\epsilon}{n^{\ast}}} \right)^{n^{\ast}} \\
= \left[ \left( \widehat{U}_{arbitrary}\right)^{\frac{1}{n^{\ast}}} \right] ^{n^{\ast}} & = & \widehat{U}_{arbitrary}.
\end{eqnarray*}

To conclude this section, let us point out that another equivalent form of
the nonholonomic control method exists,
in which the durations of the pulses are fixed, whereas the strengths of the perturbations play the role of control parameters. The equations in that case are very similar to those we have just dealt with, and the same method applies with almost no change. For more details, see \cite{HA99}.

\section{Application: implementation of a CNOT gate in a two cold Cs atom system}

\begin{figure}
[ptb]
\begin{center}
\includegraphics[
height=1.689in,
width=2.4388in
]
{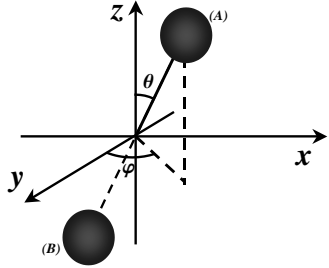}
\caption{Polar angles $\theta, \varphi$.}
\label{Fig3}
\end{center}
\end{figure}

In this section, we present the application of our control technique to a system of cold Cs atoms. Frozen Rydberg gases of interacting Cs atoms have been investigated in \cite{MCTF98} and have revealed the existence of new phenomena typical of low temperatures, such as the modification of resonance profiles, the explanation of which requires the framework of a $N$ body theory. The system we have chosen to consider in this paper is greatly inspired by the experimental situation studied in \cite{MCTF98}: it consists in two Cs atoms in Rydberg states, denoted by $(A)$ and $(B)$, of dipole momenta $\hat{\overrightarrow{d_{A}}}$ and $\hat{\overrightarrow{d_{B}}}$, respectively, linked by the fixed vector $\overrightarrow{R}=R \overrightarrow{n}$ which is determined by its norm, taken equal to $R=2 \cdot 10^{-7}m$ (of the same order as the distance between two close neighbour atoms in \cite{MCTF98}),
and its direction $\overrightarrow{n}$, defined by polar angles $\theta$ and $\varphi$ (cf Fig.\ref{Fig3}). These two atoms are coupled by dipole-dipole interaction
\[
\widehat{V}_{dd}=\frac{1}{4\pi\varepsilon_{0}R^{3}}\left[  \hat{\overrightarrow{d}}_{A} \cdot                 \hat{\overrightarrow{d}}_{B}-3\left(\hat{\overrightarrow{d}}_{A} \cdot \overrightarrow
{n}\right) \left( \hat{\overrightarrow{d}}_{B} \cdot \overrightarrow{n}\right)\right]
\] 
and are subject to a Stark field $\overrightarrow{E}_{S}=E_{S}\overrightarrow{e}_{z}$, the $z$ axis corresponding to the quantization axis for the total angular momentum. The total Hamiltonian of the system is thus composed of the unperturbed part, $\widehat{H}_{0}=\widehat{V}_{dd}$, and the controllable perturbation $\widehat{V}_{S}=-\hat{\overrightarrow{d}}\cdot \overrightarrow{E}_{S}$.  

At this stage, we shall make two remarks. Firstly, for the dipole-dipole approximation of the interaction energy to hold, \emph{i.e.} for higher order terms to be actually negligible, the distance $R$ between atoms must be much greater than the sum of the radii of the atoms, which is not strictly the case here: indeed, in the atomic Rydberg states we shall consider ($n=23,24$), the two atoms have almost the same radius, approximately equal to $3 \cdot 10^{-8}m$, whence the prescription $R \gg r_{A}+r_{B}$ is not rigorously checked. To circumvent this difficulty, one could be tempted to increase the value of the interatomic distance $R$, but this would result in a decrease of the typical value of the dipole-dipole interaction which would then become much smaller than the typical value of the Stark interaction energy: this has to be avoided for our purpose of control, and the balance between the unperturbed Hamiltonian $\widehat{H}_{0} = \widehat{V}_{dd}$ and the perturbation $\widehat{V}_{S}=-\hat{\overrightarrow{d}}\cdot \overrightarrow{E}_{S}$ must be preserved. One might then suggest to decrease the Stark field in order to make the Stark interaction term decrease and follow the dipole-dipole interaction, but then the range of the required values for the Stark field would not be realistic. We choose instead to keep the value of $R=2 \cdot 10^{-7}m$ and to consider the dipole-dipole term only, bearing in mind that a more rigorous approach should take higher order effects into account. 
Secondly, contrary to the experimental situation described in \cite{MCTF98}, we do not deal with a sample of $N$ interacting Cs atoms but only with a two atom system. Nevertheless, it has been demonstrated that this situation could be experimentally achieved \cite{SRPG01}.

Let us now describe the control operation we want to achieve. Initially in zero field, the system is prepared in an arbitrary superposition $\sum_{i=0}^{3}c_{i}\left|i\right\rangle$ of the following four states
\begin{eqnarray*}
\left|  0\right\rangle  & = & \left|  24s_{1/2},m_{j}=1/2;23s_{1/2},m_{j}=1/2\right\rangle \label{base1}\\
\left|  1\right\rangle  & = & \left|  23p_{3/2},m_{j}=3/2;23p_{3/2},m_{j}=3/2\right\rangle \label{base2}\\
\left|  2\right\rangle  & = & \left|  23p_{3/2},m_{j}=3/2;23p_{3/2},m_{j}=1/2\right\rangle \label{base3}\\
\left|  3\right\rangle  & = & \left|  23p_{3/2},m_{j}=1/2;23p_{3/2},m_{j}=1/2\right\rangle \label{base4}
\end{eqnarray*}
which formally stands for two qubits of information to be processed. Our goal is to impose the $CNOT$ gate, or, in other words, whichever the initial state
\[
\left|  \varphi\left(  0\right)  \right\rangle =c_{0}\left|  0\right\rangle
+c_{1}\left|  1\right\rangle +c_{2}\left|  2\right\rangle +c_{3}\left|
3\right\rangle
\]
is, we want to impose the particular evolution $CNOT$, yielding the final state
\[
CNOT\left|  \varphi\left(  0\right)
\right\rangle =c_{0}\left|  0\right\rangle +c_{1}\left|  1\right\rangle
+c_{3}\left|  2\right\rangle +c_{2}\left|  3\right\rangle .
\]
Let us underline that the choice of this specific gate is purely arbitrary: we could have chosen any other unitary evolution of the system. Nevertheless, the $CNOT$ gate is particularly important since it enters into the composition of universal sets of quantum gates, as demonstrated by D.P. DiVincenzo in his well-known paper \cite{DiV95}. 

\begin{figure}
[ptb]
\begin{center}
\includegraphics[height=2 in,width= 2.8 in]{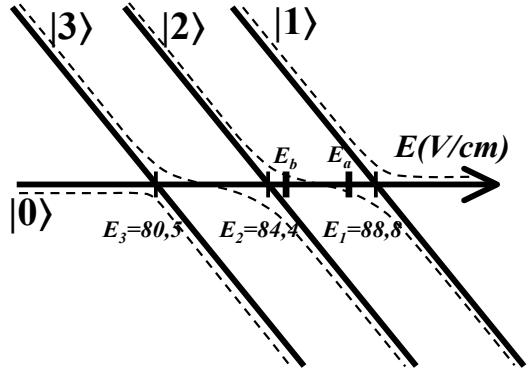}
\caption{Simplified Stark diagram of the system considered: with (dashed line) and without $V_{dd}$ (solid line).}
\label{Fig4}
\end{center}
\end{figure}

The control sequence which allows us to achieve this objective roughly consists in alternately and diabatically (\emph{i.e.} abruptly) applying the two different values $E_{a}=87.42V/cm$ and $E_{b}=84.85V/cm$
of the Stark field to the system, corresponding to two different values $\widehat{H}_{a}$ and $\widehat{H}_{b}$ of the total Hamiltonian, during pulses the durations of which will be determined through the algorithm described in the previous section. From now on, we shall make the two following physical assumptions on the system: (\emph{i}) the states $\left|  23s_{1/2},m_{j} = 1/2\right\rangle$ and $\left|  24s_{1/2},m_{j} = 1/2\right\rangle$ are not mixed with the Stark multiplicities $n=19,20$ (this assumption is motivated by the position of the avoided crossings, which are far from the values $E_{a}$ and $E_{b}$ of the applied field); (\emph{ii}) the states $\left|  23p_{3/2},m_{j} = 1/2\right\rangle$ and $\left|  23p_{3/2},m_{j} = 3/2\right\rangle$ remain unaltered (we neglect their mixing with $22d$ states, which is correct up to $10 \%$), while their energies decrease linearly with the amplitude of the applied field. 
According to these simplifications, the spectrum of our system in a static electric field can be represented as shown on Fig.\ref{Fig4}: the energies of the states $\left|1\right\rangle,\left|2\right\rangle$
and $\left|3\right\rangle$ vary linearly with the applied field (with the same slope $\gamma = -283.044$ atomic units), while the energy of the state $\left|  0\right\rangle$ remains constant; moreover, for the resonance fields $E_{1}=88.8V/cm=1.73 \cdot 10^{-8}a.u.,E_{2}=84.4V/cm=1.64 \cdot 10^{-8}a.u.,E_{3}=80.5V/cm=1.57 \cdot 10^{-8}a.u.$, we have $E\left( \left|  1\right\rangle \right) = E\left( \left|  0\right\rangle \right)$, $E\left( \left|  2\right\rangle \right) = E\left( \left|  0\right\rangle \right)$ and $E\left( \left|  3\right\rangle \right) = E\left( \left|  0\right\rangle \right)$, respectively.   
In the basis $\left\{  \left|  0\right\rangle ,\left|  1\right\rangle ,\left|
2\right\rangle ,\left|  3\right\rangle \right\}$, the total Hamiltonian takes thus the following expression
\[
\widehat{H}=\widehat{V}_{dd}+\left(
\begin{array}
[c]{cccc}%
0 & 0 & 0 & 0\\
0 & \gamma(E-E_{1}) & 0 & 0\\
0 & 0 & \gamma(E-E_{2}) & 0\\
0 & 0 & 0 & \gamma(E-E_{3})
\end{array}
\right)  , 
\] 
which will alternately take the two distinct values $\widehat{H}_{a}$ for $E=E_{a}$ and $\widehat{H}_{b}$ for $E=E_{b}$. 

In summary, we deal with a system whose Hilbert space is restricted to $N=4$ states, to which we want to apply the $CNOT$ gate. To this end, we propose to alternately apply two Hamiltonians $\widehat{H}_{a}$ and $\widehat{H}_{b}$ during pulses whose timings $\left\{ \tau_{k} \right\}$ are to be determined by the method we presented in the previous section. What we have to do first is to find the $N=4$ timings $\left\{ T_{k=1,\ldots,4} \right\}$ which meet Eq.(\ref{RacNId}) by minimizing $\sum_{j=0}^{4}\left|  a_{j}\left(  \left\{ T_{k=1,\ldots,4} \right\}
\right)  \right|  ^{2}$. Using Eq.(\ref{temps}) we then build the $N^{2}=16$-dimensional time vector $\overrightarrow{\tau}^{(0)}$ from the $T_{k}$'s, which achieves the evolution $\widehat{I}$. Then we apply the iterative algorithm described in the previous section. As the $CNOT$ gate is far from the identity matrix, we have to divide the work: we take $CNOT^{\frac{1}{n}}$ as our target evolution, where $n$ is an integer greater than $1$ for which our algorithm converges, providing the time vector $\overrightarrow{\tau}^{\frac{1}{n}}$ ; then we take the new target $CNOT^{\frac{1}{n'}}$ where $n'<n$ and run our algorithm again, yielding the vector $\overrightarrow{\tau}^{\frac{1}{n'}}$, etc. as long as we obtain convergence. The smallest value of $n$ we obtained is $n^{*}=8 $, associated with the time vector $\overrightarrow{\tau}^{\frac{1}{n^{*}}}$ which achieves the evolution $CNOT^{\frac{1}{n^{*}}}$: the desired evolution $CNOT$ is obtained by repeating $n^{*}=8 $ times the same elementary control sequence, defined by $\overrightarrow{\tau}^{\frac{1}{n^{*}}}$. We thus see that the time vector $\overrightarrow{\tau}^{CNOT}$ which achieves $CNOT$ is $n^{*} \times N^{2} = 8 \times 16 =128$-dimensional and can be built by repeating $8$ times the vector $\overrightarrow{\tau}^{\frac{1}{n^{*}}}$.     

\begin{figure}
[ptb]
\begin{center}
\includegraphics[
height=2.3in,
width=3.2in
]
{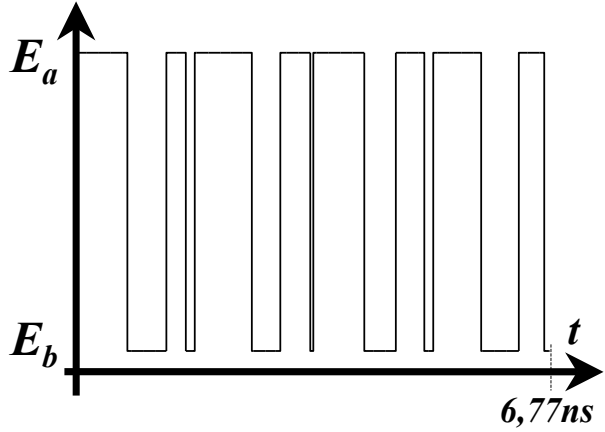}
\caption{Elementary $N^{2}=16$ pulse control sequence obtained for $R=2 \cdot 10^{-7}m, \theta=\pi/15$, and $\phi=\pi/6$. The total duration is $6,77ns$ and the values of the different interaction times are $t_{i}(ns)=\{$0.743378, 0.553823, 0.277301, 0.133699, 0.800748, 0.423586, 0.427981, 0.0427037, 0.71635, 0.458626, 0.403841, 0.13241, 0.682599, 0.54579, 0.349389, 0.0809227$\}$.}
\label{Fig5}
\end{center}
\end{figure}

Fig.\ref{Fig5} presents the numerical results we obtained for $\theta=\pi / 15$ and $\phi = \pi / 6$ (cf Fig.\ref{Fig3}) which shows the switchings of the Stark field on an elementary control sequence, whose duration is $6.77ns$. In our calculations, we tried to remain in a realistic range for the different parameters of our system. For example, the total control duration, which is of the order of
$8 \times 6.77ns \simeq 0.06\mu s$, is much smaller than the lifetime of the Rydberg states considered, which is approximately $10\mu s$. Yet, serious problems and limitations arise.

Firstly, the required switching time of the Stark field is much less than $1 ns$ (some timings are a few $100ps$), which is experimentally very difficult to achieve: this will unavoidably threaten the reliability of the control. To address this problem, one may consider replacing the static control fields by pulsed lasers, as in \cite{BACDHKKMP05}, which would probably allow rapid switching times and would certainly be more tractable experimentally. Secondly, the four state system we have considered here is a severe idealization: the couplings between the states
\begin{eqnarray*}
&  \left|  24s_{1/2},m_{j}=\pm 1/2;23s_{1/2},m_{j}=\pm 1/2\right\rangle \\
&  \left|  23p_{3/2},m_{j}=\pm3/2;23p_{3/2},m_{j}=\pm3/2\right\rangle \\
&  \left|  23p_{3/2},m_{j}=\pm3/2;23p_{3/2},m_{j}=\pm1/2\right\rangle \\
&  \left|  23p_{3/2},m_{j}=\pm1/2;23p_{3/2},m_{j}=\pm1/2\right\rangle 
\end{eqnarray*}
cannot be ignored. In addition, the influence on the states $23s$ and $24s$ of the multiplicities $n=19,20$ has been completely neglected, as well as the mixing of the states $23p_{3/2}$ with the states $22d$. One can solve these problems by increasing the state space, \emph{i.e.} by taking all the states which are actually coupled by the Stark field and the dipole-dipole interaction into account: calculating the control time vector becomes much longer, as the system considered is larger, but, fundamentally, the structure of the problem remains the same. Finally, if we do not work with two atoms but rather with a large sample, it might be experimentally very difficult to fix precise values to $R,\theta$ and $\varphi$:
this results again in a loss of reliability of the control. A possible solution to this problem, though not perfect, would be to put the atoms in an optical lattice, which would allow one to control more precisely their spatial arrangement. Another method would be to perform a first control sequence, the goal of which would be to distinguish between ''good'' and ''bad'' pairs of atoms: for instance, starting from the state $\left|3\right\rangle$, the ''good'' pairs (\emph{i.e.} the pairs with the required vector $\overrightarrow{R}$) will undergo the CNOT gate and will thus end in the state $\left|4\right\rangle$, whereas the other pairs will end in a superposition of all states and could therefore be experimentally distinguished and destroyed. 

To conclude this section, we want to emphasize that the limitations discussed
above do not remove the pedagogical and demonstrative value of the application presented.
The example considered here shows the operability of the nonholonomic control
method and suggests that it can be actually employed to achieve real objectives of control.

\section{Conclusion}
In this paper, we first recalled the nonholonomic control technique,
which allows one to control the evolution operator of generic quantum systems which
meet the bracket generation condition: after putting it in the general framework of quantum control we briefly exposed its main algorithmic features and underlined the fundamental reasons for its convergence. Then we suggested a demonstrative application of this scheme to a system of two cold Cs atoms, inspired by experimental studies on cold Rydberg gases: we showed that through alternately applying two different values of a Stark field during $128$ pulses, the timings of which range from $40ps$ to $800ps$, one can impose the $CNOT$ gate to two qubits of information stored in four specific states of the system. Finally we discussed the physical validity and the limitations of our application.      

\bigskip
The authors thank V.M. Akulin and P. Pillet (Laboratoire Aim\'{e} Cotton, Orsay, France) for stimulating and fruitful discussions. The support of EU (QUACS RTN) is kindly acknowledged.

\end{document}